\documentclass{emulateapj}
\usepackage{xspace}
\usepackage{amsmath}
\usepackage{hyperref}
\usepackage{framed} 
\usepackage{txfonts}

\def\mcut{M_{\rm cut}}
\def\lcut{L_{\rm cut}}
\def\muv{M_{1500}}
\def\HII{{\rm HII}}
\def\qi{Q_\HII}

\def\dim#1{\mbox{\,#1}}
\def\hide#1{}

\begin{document}

\title{Cosmic Reionization On Computers: The Faint End of the Galaxy Luminosity Function}

\author{Nickolay Y.\ Gnedin\altaffilmark{1,2,3}}
\altaffiltext{1}{Particle Astrophysics Center, Fermi National Accelerator Laboratory, Batavia, IL 60510, USA; gnedin@fnal.gov}
\altaffiltext{2}{Kavli Institute for Cosmological Physics, The University of Chicago, Chicago, IL 60637 USA;}
\altaffiltext{3}{Department of Astronomy \& Astrophysics, The
  University of Chicago, Chicago, IL 60637 USA} 

\begin{abstract}
Using numerical cosmological simulations completed under the ``Cosmic Reionization On Computers'' (CROC) project, I explore theoretical predictions for the faint end of the galaxy UV luminosity functions at $z\ga6$. A commonly used Schechter function approximation with the magnitude cut at $\mcut\sim-13$ provides a reasonable fit to the actual luminosity function of simulated galaxies. When the Schechter functional form is forced on the luminosity functions from the simulations, the magnitude cut $\mcut$ is found to vary between $-12$ and $-14$ with a mild redshift dependence. An analytical model of reionization from Madau, Haardt \& Rees (1997), as used by Robertson et al.\ (2015), provides a good description of the simulated results, which can be improved even further by adding two physically motivated modifications to the original Madau, Haardt \& Rees (1997) equation.
\end{abstract}


\section{Introduction}
\label{sec:intro}

It is well established by now that dwarf galaxies contribute significantly to the overall budget of ionizing photons during cosmic reionization, and that the slope of the Schechter function fit to the observational measurements of galaxy UV luminosity functions can approach and even exceed -2
 \citep{gals:biol11,rei:obig12,rei:btos12,rei:sreo13,rei:wmhb13,rei:obil13,rei:obi14,rei:wjc14,rei:bdm14,rei:refd15,rei:ark15,rei:sfa15,rei:frp15,rei:arj15}. At the slope value of -2 the total luminosity density of the Schechter fit diverges, so the Schechter function can remain a valid approximation to the actual galaxy luminosity function only over a limited range of luminosities. A commonly used resolution of this difficulty is to introduce a cutoff luminosity or magnitude $\mcut$ to the Schechter approximation \citep{rei:rfsc13,rei:refd15}. 
 
Such a cutoff is necessarily arbitrary, and will remain so for the foreseeable future, as the actual values of $\mcut \sim -13$ are not only beyond the limit of the existing HST observations, but will also be hard to reach even by JWST observations of lensed reionization sources.
 
Therefore, it may make sense to explore what the current theory has to say on the specific shape of the galaxy UV luminosity function and possible reasonable values of $\mcut$. In this paper I use numerical simulation of reionization completed under the ``Cosmic Reionization On Computers'' (CROC) project as a theoretical tool. CROC simulations are a suitable tool for this purpose, since they, currently, match all existing observational constraints, from galactic properties such as luminosity functions, UV slopes, and IR excesses, to the properties of the IGM at $z<6$ such as full PDFs of Gunn-Peterson optical depth in several redshift bins, flux gap statistics, etc \citep{ng:gk14,ng:lg16,ng:gbf16}. They are also comparable to or exceeding other modern, fully self-consistent simulations of reionization in mass and spatial resolution and in the box size. 

CROC simulations are also useful for modeling galaxy UV luminosity functions, since their latest series has been shown to provide numerically (albeit weakly) converged results \citep{ng:g16a}. Hence, numerical effects are under control in CROC simulations and do not exceed 20\% for galaxy luminosity functions at all redshifts.

The full details of numerical setups of CROC simulations are presented in \citet{ng:g14} and \citet{ng:g16a}, and I do not repeat them here for the sake of brevity. In this paper I use a new ``Cayman'' series of simulations that maintain spatial resolution fixed in physical units (rather than in comoving units, as the first generation of CROC runs) and include weak convergence corrections from \citet{ng:g16a} that compensate numerical results for finite spatial and mass resolution and make then approximately resolution independent (and equal to the fully numerically converged values). In the rest of the paper simulation sets are labeled in the following way: the label starts with the box size (B20 standing for the $20h^{-1}\dim{Mpc}$ box size) followed by the mass resolution (MR for ``medium'' and HR for ``high'' mass resolution, which stand for $512^3$ and $1024^3$ initial grids in $20h^{-1}\dim{Mpc}$ boxes and proportionally larger grids in larger boxes respectively) and concluded with the spatial resolution (R100 standing for the spatial resolution of $100\dim{pc}$ in physical units).

\section{The Faint End of the Galaxy UV Luminosity Function}

\begin{figure*}[t]
\includegraphics[width=0.5\hsize]{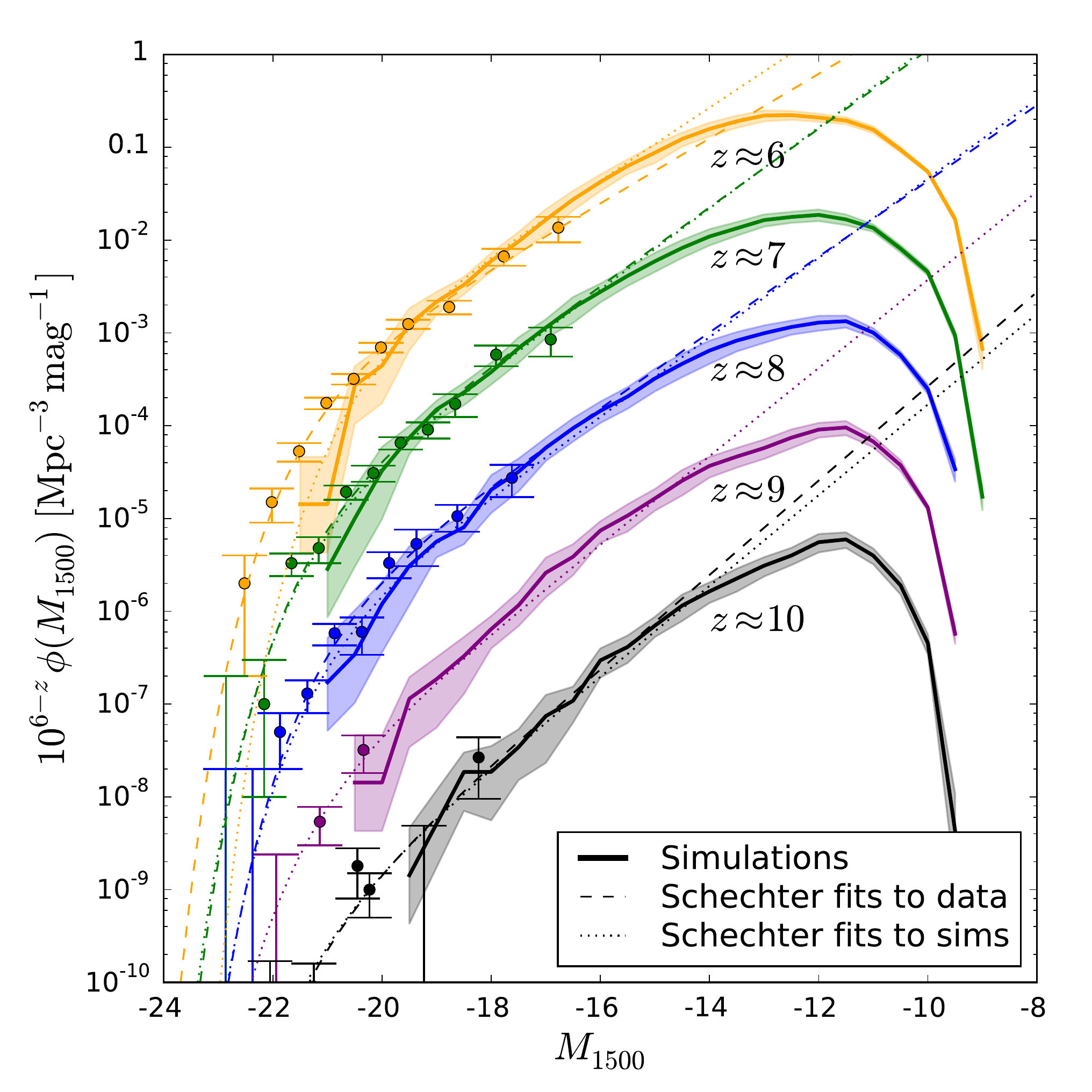}%
\includegraphics[width=0.5\hsize]{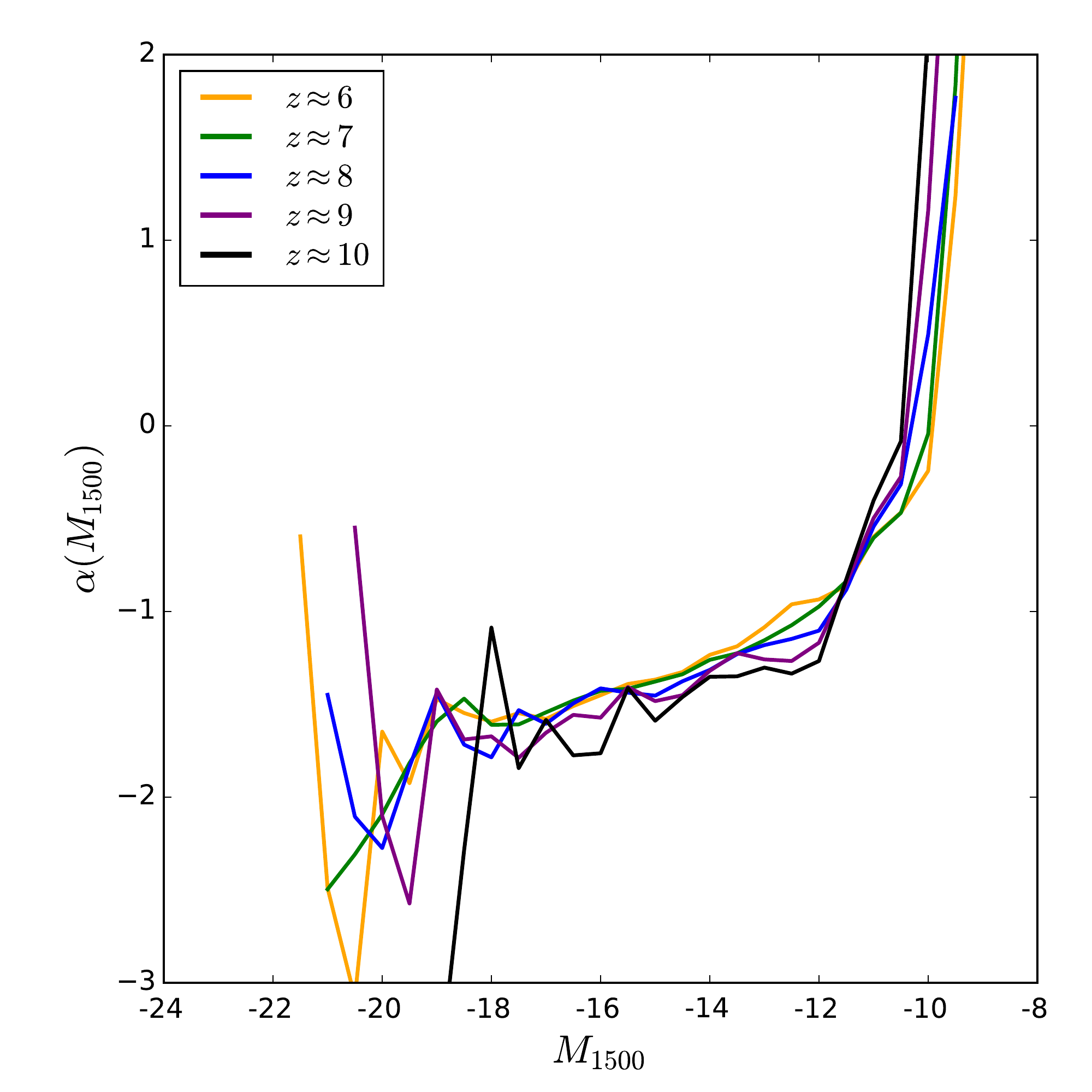}%
\caption{Galaxy UV luminosity functions (left) and their local slopes (right) at 5 different redshifts for the fiducial B20MR.R100 simulation set. Points with error-bars and dotted lines in the left panel are observational data and their Schechter function fits from \protect\citet{rei:biot15}, while the dashed lines are Schechter function fits to the simulation results in the interval $-22<M_{1500}<-16$. \label{fig:lfdif}}
\end{figure*}

Galaxy UV luminosity functions at a range of redshifts are a primary reionization observable, and provide stringent constraints on the properties of reionization sources. In Figure \ref{fig:lfdif} I shows galaxy luminosity functions at $z\ga6$ and their logarithmic slopes, as well as their Schechter function fits, for the fiducial set of 6 independent realizations of the B20MR.R100 box. Simulated luminosity functions peak at $\muv\sim-12$ and drop rapidly at lower magnitudes. The specific shape of this drop is \emph{not} a reliable prediction of the simulations - it depends on the details of the star formation algorithm, and, in particular, on the adopted minimum mass of a stellar particle. These details, however, do not affect the total luminosity density or the shape of the luminosity function at $\muv<-12$.

\begin{figure}[b]
\includegraphics[width=\hsize]{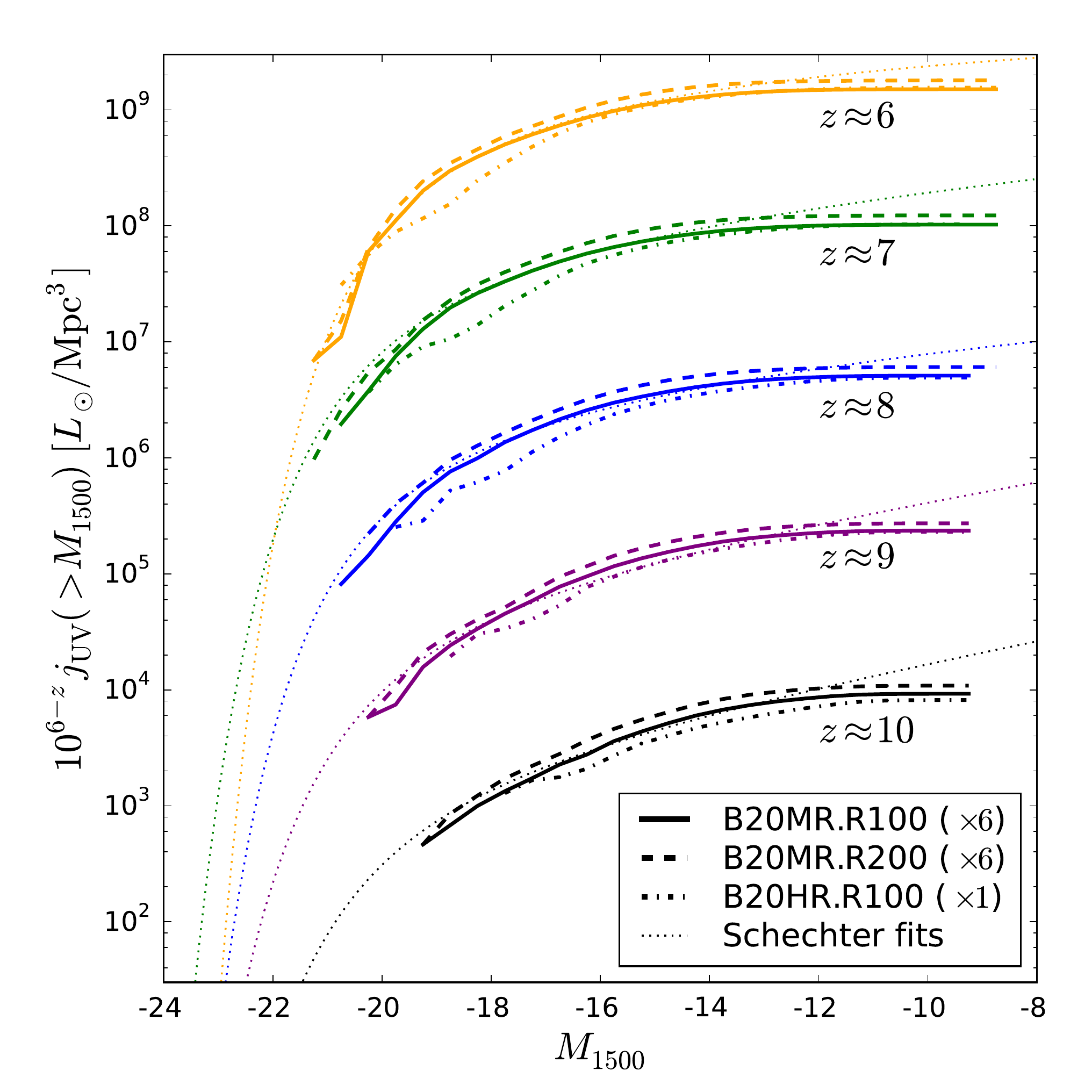}%
\caption{Cumulative galaxy UV luminosity functions for 3 different simulation sets with different mass and spatial resolution and Schechter function fits to the fiducial set B20MR.R100. \label{fig:lfcum}}
\end{figure}

A parameter, which is of a particular interest to analytical modeling, is the limiting magnitude $\mcut$ to which the Schechter fit to the luminosity function must be integrated to recover the total luminosity density. To illustrate its role, I show in Figure \ref{fig:lfcum} cumulative luminosity functions for 3 different simulation sets with varied spatial and mass resolution. Actual model luminosity functions start deviating from their Schechter fits at $\muv\sim-14$, and reach their asymptotic values at $\muv>-12$. One can then define the cutoff luminosity $\lcut$ for the Schechter function fit so that the total luminosity density $j_{\rm UV}$ in the actual simulated luminosity function and its Schechter fit are equal,
\begin{equation}
  j_{\rm UV} \equiv \int_0^\infty \phi_{\rm SIM}(L)\, L\, dL = \int_{L_{\rm cut}}^\infty \phi_{\rm SCH}(L)\, L\, dL.
  \label{eq:juv}
\end{equation}

\begin{figure}[t]
\includegraphics[width=\hsize]{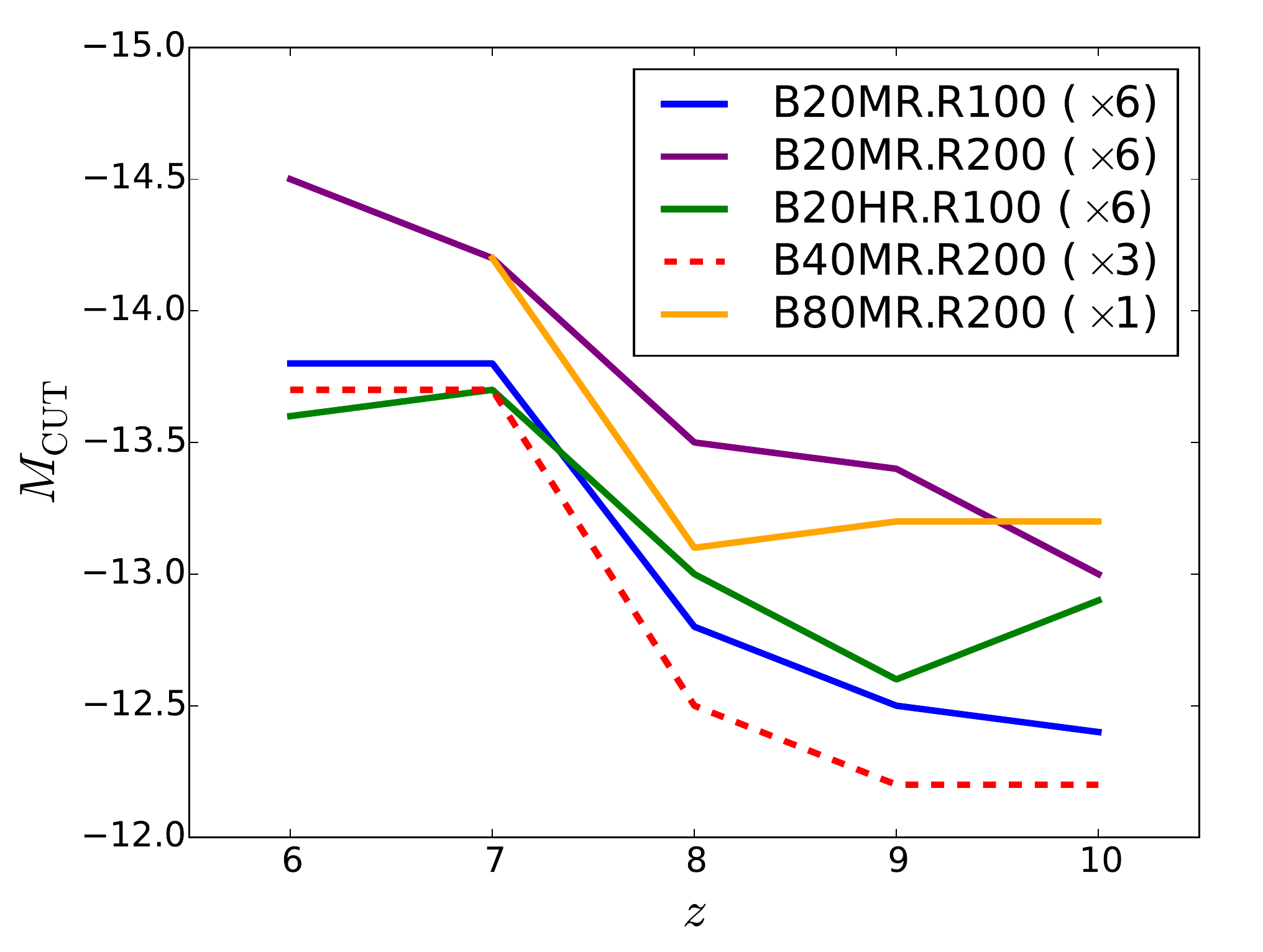}%
\caption{Cutoff magnitudes $\mcut$ to which Schechter function fits need to be integrated to recover the correct total UV luminosity density, as functions of redshift, for several simulation sets. \label{fig:magcut}}
\end{figure}

Corresponding magnitude cuts are plotted in Figure \ref{fig:magcut} for the three simulation sets used above, and two larger box sets (to test the effect of the box size). About 1 magnitude difference between different simulations sets should be treated as the estimate of the theoretical uncertainty, as all sets are weakly numerically converged \citep{ng:g16a}; however, some modest, below 20\%, residual dependence on the mass and spatial resolution and on the box size remains. The cutoff magnitude is slightly redshift dependent, but that dependence is too mild to significantly affect ionization history modeling discussed below.

\section{Ionization History Modeling}

Galaxy UV luminosity functions are often used in modeling reionization history of the universe. The simplest form of such modeling was introduced by \citet{igm:mhr99}; it is based on a single evolution equation for the filling factor of ionized gas $\qi$,
\begin{equation}
  \frac{d\qi}{dt} = \frac{\dot{n}_{\rm ion}}{n_{\rm H}} - \frac{\qi}{\bar{t}_{\rm rec}},
  \label{eq:mhr}
\end{equation}
where $\dot{n}_{\rm ion}$ is the globally averaged rate of production of hydrogen ionizing photons, $n_{\rm H}$ is the averaged hydrogen nuclei density, and $\bar{t}_{\rm rec}$ is the harmonically averaged, ionizing gas mass-weighted hydrogen recombination time,
\begin{equation}
  \bar{t}_{\rm rec} \equiv \langle x_i/t_{\rm rec}\rangle_M/\langle x_i\rangle_M.
  \label{eq:trec}
\end{equation}
\citet{igm:mhr99} type modeling remains a useful tool despite its simplicity; in particular, I use hereafter the work of \citet{rei:refd15} as one of the most recent and widely regarded analytical models of reionization.

Equation \ref{eq:mhr} has one major limitation: it assumes that all photons are expended on ionizing general IGM, hence ignoring ionizing photon loss in Lyman limit systems \citep{reisam:fo05,igm:fm09,ng:kg13}, which becomes important closer to the end of reionization.

The term $\dot{n}_{\rm ion}$ is commonly evaluated as \citep{rei:refd15}
\[
  \dot{n}_{\rm ion} = f_{\rm esc}^{\rm eff} \dot{n}_{\rm emit},
\]
where $ f^{\rm eff}_{\rm esc}$ is the ``effective'' escape fraction of ionizing photons and $\dot{n}_{\rm emit}$ is the rate of emission of ionizing photons by stellar sources,
\begin{equation}
  \dot{n}_{\rm emit} = \xi_{\rm ion} \dot{\rho}_* = \xi_{\rm ion} \kappa_{\rm UV} j_{\rm UV}.
  \label{eq:nion}
\end{equation}
Here $\xi_{\rm ion}$ is the ionizing photon production efficiency per unit star formation rate, $\kappa_{\rm UV}$ is the conversion factor from UV luminosity to star formation rate, and $j_{\rm UV}$ is the UV luminosity density from equation (\ref{eq:juv}).

\begin{figure}[t]
\includegraphics[width=\hsize]{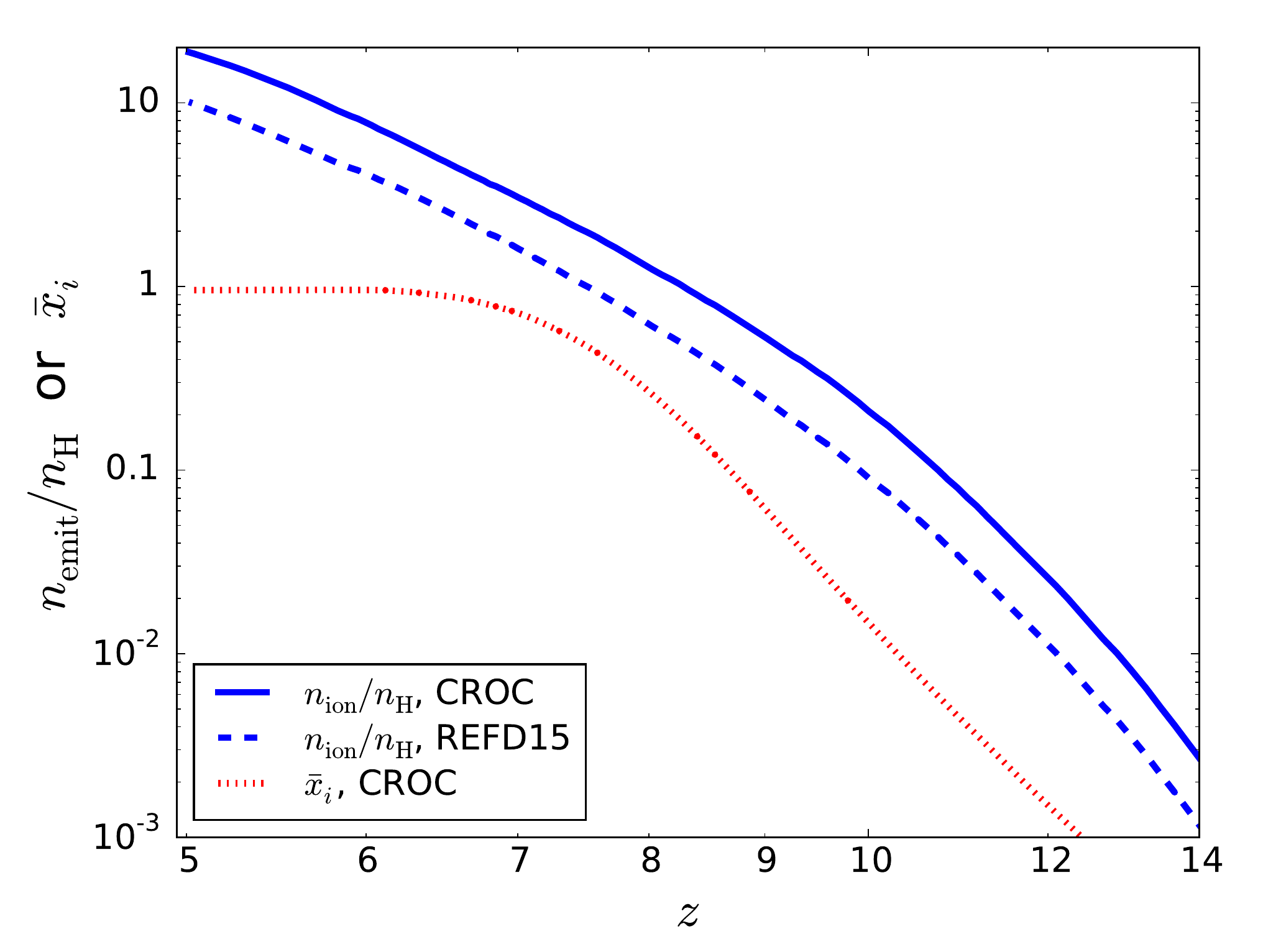}%
\caption{Evolution of the number of emitted (i.e.\ before accounting for the escape fraction) ionizing photons per hydrogen nucleus, as actually produced in the simulations (solid blue line) and as computed from the globally averaged start formation rate (dashed blue line) using equation (\ref{eq:nion}) with the choice of parameters from \protect\citet{rei:refd15}. The dotted red line shows the mass weighted average neutral fraction $\langle x_i\rangle_M$.\label{fig:nion}}
\end{figure}

To explore the connection with the analytical modeling even further, I show in Figure \ref{fig:nion} both the production rate of ionizing photons per hydrogen nucleus $\dot{n}_{\rm emit}/n_{\rm H}$ and the actual mass-weighted ionized fraction $\langle x_i\rangle_M$ - the specific values of other parameters are set exactly as in \citet{rei:refd15}. 

The ionizing photon production rate can be computed in two ways: either extracted directly from the simulations or converted from the global star formation rate via equation (\ref{eq:nion}). Both approaches give values that differ up to a factor of 2, and the difference is primarily due to the factor $\xi_{\rm ion}$ - while \citet{rei:refd15} assume a fixed value for this quantity, in the CROC simulations each stellar particle uses its own, metallicity-dependent factor $\xi_{\rm ion}$, as computed by Startburst99 code \citep[see][for details]{ng:g14}. The CROC value is in better agreement with the value of $\xi_{\rm ion}$ from \citet{gals:ts15}.

\begin{figure}[t]
\includegraphics[width=\hsize]{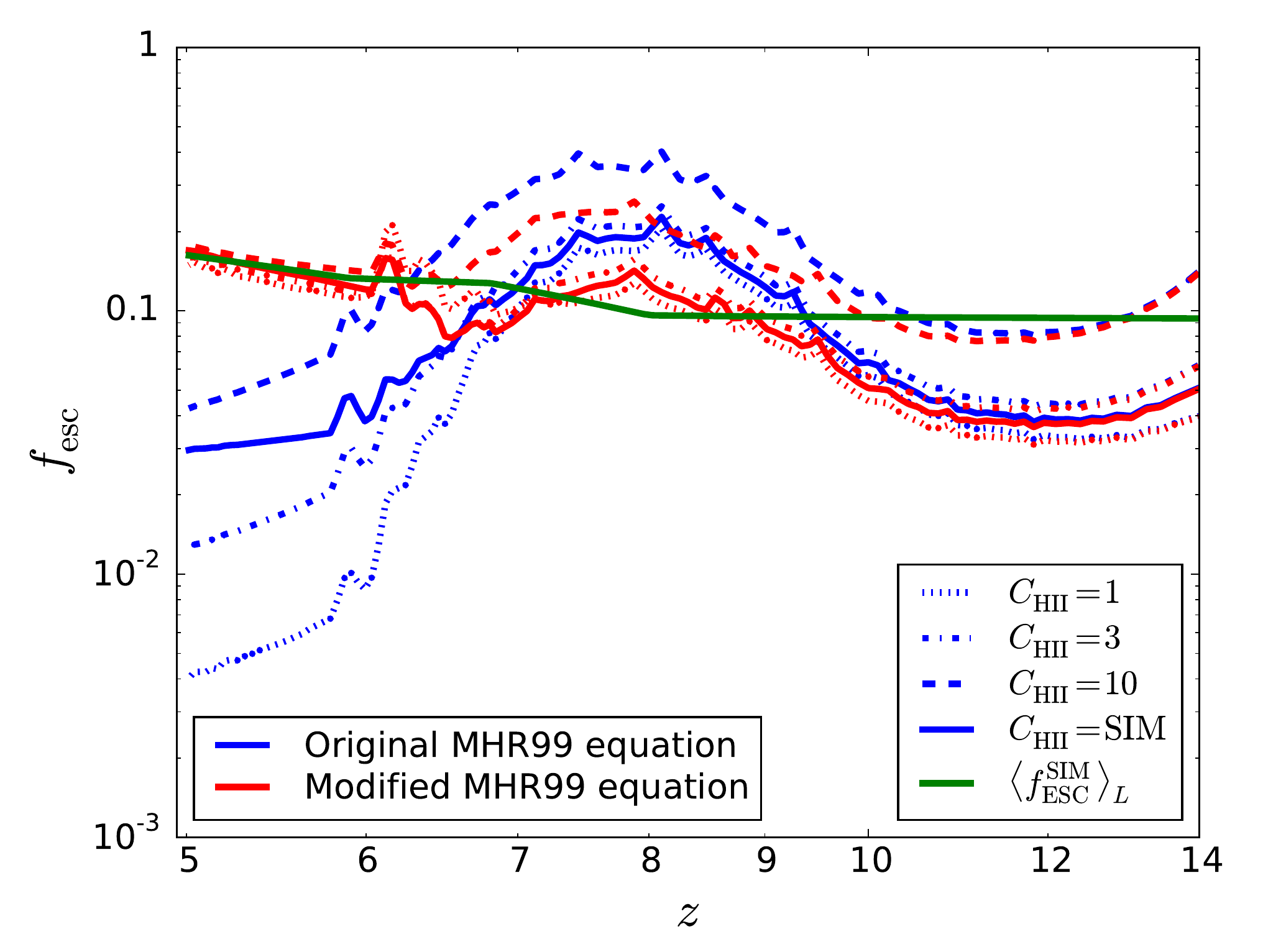}%
\caption{Effective escape fraction at several values of the assumed clumping factor in the \protect\citet{rei:refd15} model for the original \protect\citet{igm:mhr99} equation \ref{eq:mhr} (blue lines) and for the modified \protect\citet{igm:mhr99} equation \ref{eq:mhrmod} (red lines). Solid lines assume the clumping factor from the actual simulations \protect\citet[Fig.\ 6 from][]{ng:kg15}. The solid green function tracts the luminosity-weighted average escape fraction estimated from the actual simulations.\label{fig:fesc}}
\end{figure}

Given $\dot{n}_{\rm emit}$, the clumping factor of the ionized gas $C_{\HII}$ that enters the average recombination $\bar{t}_{\rm rec}$ as functions of time, and adopting a commonly used approximation $\qi \approx \langle x_i\rangle_M$, one can compute the effective escape fraction from equation \ref{eq:mhr}. The results of such computation are shown in Figure \ref{fig:fesc} with blue lines for four different models of the clumping factor $C_{\HII}$. For comparison, the green line shows the actual luminosity weighted average escape fraction from the simulations. While the latter vary little with time, the effective escape fraction $f_{\rm esc}^{\rm eff}$ one needs to use in equation (\ref{eq:mhr}) varies in a non-trivial way, first increasing by a factor of 3 at $z\sim 8$ and later dropping rapidly by an order of magnitude or more. Such a complex behavior, not mirrored in the actual simulation, needs to be explained.

The rapid decrease in the effective escape fraction at $z \la 7$ is most likely due to the limitation of equation (\ref{eq:mhr}) discussed above - the lack of accounting for the Lyman Limit systems. At the end of reionization most of ionizing photons are absorbed in the Lyman Limit systems, and only a small fraction of them is used for ionizing the last remnants of the neutral IGM.

Understanding the peak in the effective escape fraction at $z\sim8$ is harder. The reason for this feature is apparent from Fig.\ \ref{fig:nion}: after $z\sim9$ the rate of increase in $\dot{n}_{\rm emit}$ slows down, while $\langle x_i\rangle_M$ continues increasing rapidly and slows down only by $z\sim 7$. Hence, to maintain the rapid increase in $\langle x_i\rangle_M$, the effective escape fraction has to increase.

One possibility that should always be considered in the simulation work is the existence of numerical artifacts. CROC simulations use a ``Reduced Speed of Light'' approximation \citep{ng:ga01} for modeling radiative transfer, and that approximation may affect the solution. In order to check for such a possibility, I completed two more simulations in which the effective speed of light in the radiative transfer solver was varied by a factor of 10, between $0.03c$ and $0.3c$. Full description of these tests will be presented elsewhere; ionization histories in these test runs are almost indistinguishable from the fiducial runs, eliminating the numerical artifacts due to the ``Reduced Speed of Light'' approximation as a cause of the non-trivial behavior of the effective escape fraction.

An increase in the effective escape fraction can also be caused by additional, originating in sources other than massive stars, ionizing photons. In order to check for such a possibility, I ran another test simulation that excluded ionizing photons from quasars, helium recombination photons, and bremsstrahlung. These additional sources of photons, as implemented in the CROC simulations, are not enough to explain away the variation in $f_{\rm esc}^{\rm eff}$.

Another reason for the delay in the evolution of $\langle x_i\rangle_M$ as compared to the rate of increase in $\dot{n}_{\rm emit}$ is the actual physical delay between the moment the photon is emitted and the time it is absorbed by a neutral IGM atom. In order to account for such a possibility (and for the Lyman limit systems as well), I consider a modified version of the \citet{igm:mhr99} equation,
\begin{equation}
  \frac{d\qi}{dt} = \frac{\dot{n}_{\rm ion}\left|_{t-t_{\rm del}}\right.}{n_{\rm H}}\left(1-f_{\rm LLS}\right) - \frac{\qi}{\bar{t}_{\rm rec}},
  \label{eq:mhrmod}
\end{equation}
where $t_{\rm del}$ is the effective delay time between the moment of emission and the moment of absorption and $f_{\rm LLS}$ is the fraction of ionizing photons lost in the Lyman limit systems. As an example, I adopt the following ansatzes for these two new factors, $t_{\rm del} = \min\left(50,600/\psi(z)\right)\dim{Myr}$ and $f_{\rm LLS} = \exp\left(-1.5\psi(z)(1-\qi)C_{\HII}\right)$, with $\psi(z)=\exp\left(z-6\right)$. 

Effective escape fractions computed with the modified equation (\ref{eq:mhrmod}) are shown in Fig.\ \ref{fig:fesc} with red lines. The two corrections eliminate most of the variation in $f_{\rm esc}^{\rm eff}$ for $z<10$ and also reduce the dependence on $C_{\HII}$. The variation in $f_{\rm esc}^{\rm eff}$ at higher redshifts is, perhaps, not surprising, as equations (\ref{eq:mhr}) and (\ref{eq:mhrmod}) may not capture the very first stages of reionization well, when ionized bubbles occupy only a small fraction of the total volume and, hence, do not sample the average universe.

Of course, these two corrections should only be considered as a mere illustration, especially since the maximum adopted value for $t_{\rm del}$ appears to be a bit too large. I made no effort in motivating the fitting functions from some a priori physical grounds, as such effort would be well outside the scope of this paper.

\section{Conclusions}
 
Galaxy UV luminosity functions, as modeled by CROC simulations, agree well with the existing observational measurements. When fitted with a Schechter functional form, they require the magnitude cut between -12 and -14, with only slight redshift dependence, in good agreement with assumptions made in the \citet{igm:mhr99} style analytical modeling using equation (\ref{eq:mhr}) \citep[c.f][as the latest example of such modeling]{rei:refd15}.

However, a more serious problem with this type of modeling is the adopted assumptions about the effective escape fraction of ionizing radiation. While actual escape fraction in the simulations are approximately constant in time and with galaxy luminosity, intrinsic limitations of equation (\ref{eq:mhr}) make the effective escape fraction one has to use in equation (\ref{eq:nion}) strongly and non-trivially redshift-dependent. One of these limitations is not accounting for photon loss in the Lyman limit systems, which causes a rapid drop in the effective escape fraction at the end of reionization.

A simple modification of equation (\ref{eq:mhr}) as given by equation  (\ref{eq:mhrmod}) can eliminate most of this variations, although the question of whether such a modification is fully warranted and physically justified remains to be answered.

\acknowledgements

I am grateful to Brant Robertson for valuable comments on the earlier draft and for the supportive referee report.

Fermilab is operated by Fermi Research Alliance, LLC, under Contract No.~DE-AC02-07CH11359 with the United States Department of Energy. This work was also supported in part by the NSF grant AST-1211190 and by the Munich Institute for Astro- and Particle Physics (MIAPP) of the DFG cluster of excellence "Origin and Structure of the Universe". CROC simulations have been performed on the University of Chicago Research Computing Center cluster ``Midway'', on National Energy Research Supercomputing Center (NERSC) supercomputers ``Cori'' and ``Edison'', and on the Argonne Leadership Computing Facility supercomputer ``Mira''. An award of computer time was provided by the Innovative and Novel Computational Impact on Theory and Experiment (INCITE) program. This research used resources of the Argonne Leadership Computing Facility, which is a DOE Office of Science User Facility supported under Contract DE-AC02-06CH11357.

\end{document}